\documentclass[twocolumn]{aa} 
\usepackage{graphicx}
\usepackage{txfonts}

\begin{document}

\title{Modeling the orbital motion of Sgr A*'s near-infrared flares}

 \author{The GRAVITY  Collaboration: M.~Baub\"ock\inst{1}, J.~Dexter\inst{1,15}, R.~Abuter\inst{8}, A.~Amorim\inst{6}, J.P.~Berger\inst{5}, H.~Bonnet\inst{8}, W.~Brandner\inst{3}, Y.~Cl\'enet\inst{2}, V.~Coud\'e~du~Foresto\inst{2}, P.T.~de~Zeeuw\inst{10,1}, G.~Duvert\inst{5}, A.~Eckart\inst{4,13}, F.~Eisenhauer\inst{1}, N.M.~F\"orster Schreiber\inst{1}, F.~Gao\inst{1}, P.~Garcia\inst{7,9}, E.~Gendron\inst{2}, R.~Genzel\inst{1,11}, O.~Gerhard\inst{1}, S.~Gillessen\inst{1}, M.~Habibi\inst{1}, X.~Haubois\inst{9}, T.~Henning\inst{3}, S.~Hippler\inst{3}, M.~Horrobin\inst{4}, A.~Jim\'enez-Rosales\inst{1}, L.~Jocou\inst{5}, P.~Kervella\inst{2}, S.~Lacour\inst{2,1}, V.~Lapeyr\`ere\inst{2}, J.-B.~Le Bouquin\inst{5}, P.~L\'ena\inst{2}, T.~Ott\inst{1}, T.~Paumard\inst{2}, K.~Perraut\inst{5}, G.~Perrin\inst{2}, O.~Pfuhl\inst{1}, S.~Rabien\inst{1}, G.~Rodriguez Coira\inst{2}, G.~Rousset\inst{2}, S.~Scheithauer\inst{3}, J.~Stadler\inst{1}, A.~Sternberg\inst{12,14}, O.~Straub\inst{1}, C.~Straubmeier\inst{4}, E.~Sturm\inst{1}, L.J.~Tacconi\inst{1}, F.~Vincent\inst{2}, S.~von Fellenberg\inst{1}, I.~Waisberg\inst{16}, F.~Widmann\inst{1}, E.~Wieprecht\inst{1}, E.~Wiezorrek\inst{1}, J.~Woillez\inst{8} \& S.~Yazici\inst{1,4}
  \fnmsep\thanks{GRAVITY has been developed by a collaboration of the Max Planck Institute for Extraterrestrial Physics, LESIA of Paris Observatory / CNRS / UPMC / Univ. Paris Diderot and IPAG of Universit\'e Grenoble Alpes / CNRS, the Max Planck Institute for Astronomy, the University of Cologne, the Centro de Astrof\'isica e Gravita{\c{c}}\~ao, and the European Southern Observatory.\newline Corresponding author:
    M.~Baub\"ock (bauboeck@mpe.mpg.de),}
         }

\titlerunning{Orbital Motion of Sgr A*'s Flares}
\authorrunning{GRAVITY Collaboration}
\institute{Max Planck Institute for Extraterrestrial Physics (MPE), Giessenbachstr.1, 85748 Garching, Germany
\and
LESIA, Observatoire de Paris, Universit\'e PSL, CNRS, Sorbonne Universit\'e, Univ. Paris Diderot, Sorbonne Paris Cit\'e, 5 place Jules Janssen, 92195 Meudon, France
\and
Max-Planck-Institute for Astronomy, K\"onigstuhl 17, 69117 Heidelberg, Germany
\and
1. Physikalisches Institut, Universit\"at zu K\"oln, Z\"ulpicher Str. 77, 50937 K\"oln, Germany
\and
Univ. Grenoble Alpes, CNRS, IPAG, 38000 Grenoble, France
\and
CENTRA and Universidade de Lisboa - Faculdade de Ci\^encias, Campo Grande, 1749-016 Lisboa, Portugal
\and
CENTRA and Universidade do Porto - Faculdade de Engenharia, 4200-465 Porto, Portugal
\and
European Southern Observatory, Karl-Schwarzschild-Str. 2, 85748 Garching, Germany
\and
European Southern Observatory, Casilla 19001, Santiago 19, Chile
\and
Sterrewacht Leiden, Leiden University, Postbus 9513, 2300 RA Leiden, The Netherlands
\and
Departments of Physics and Astronomy, Le Conte Hall, University of California, Berkeley, CA 94720, USA
\and
School of Physics and Astronomy, Tel Aviv University, Tel Aviv 69978, Israel
\and
Max-Planck-Institute for Radio Astronomy, Auf dem H\"ugel 69, 53121 Bonn, Germany
\and
Center for Computational Astrophysics, Flatiron Institute, 162 5th Ave., New York, NY, 10010, USA
\and
JILA and Department of Astrophysical and Planetary Sciences, University of Colorado, Boulder, CO 80309, USA
\and
Department of Particle Physics and Astrophysics, Weizmann Institute of Science, Rehovot 7610001 ISRAEL
}

\abstract{Infrared observations of Sgr A* probe the region close to the event horizon of the black hole at the Galactic center. These observations can constrain the properties of low-luminosity accretion as well as that of the black hole itself. The GRAVITY instrument at the ESO VLTI has recently detected continuous circular relativistic motion during infrared flares which has been interpreted as orbital motion near the event horizon. Here we analyze the astrometric data from these flares, taking into account the effects of out-of-plane motion and orbital shear of material near the event horizon of the black hole. We have developed a new code to predict astrometric motion and flux variability from compact emission regions following particle orbits. Our code combines semi-analytic calculations of timelike geodesics that allow for out-of-plane or elliptical motions with ray tracing of photon trajectories to compute time-dependent images and light curves. We apply our code to the three flares observed with GRAVITY in 2018. We show that all flares are consistent with a hotspot orbiting at R$\sim$9 gravitational radii with an inclination of $i\sim140^\circ$. The emitting region must be compact and less than $\sim5$ gravitational radii in diameter. We place a further limit on the out-of-plane motion during the flare.}

\maketitle

\section{Introduction}
The black hole at the Galactic center provides a unique laboratory for exploring the physics of accretion and supermassive black holes. As it is the closest supermassive black hole, Sgr~A* has the largest angular size of any black hole observable from Earth, with a Schwarzschild radius of $R_s = 10.022 \pm 0.020_{\rm stat.} \pm .032_{\rm sys.} \mu \mathrm{as}$ \citep{grav19}. Additionally, the characteristic timescale of infalling matter near the black hole is on the order of minutes to hours (e.g.,\ \citealt{bag03, mey09, wit12, hor14, wit18}), allowing variability and motion to be observed within a single night.  

The GRAVITY instrument at the VLT Interferometer (\citealt{eis08, grav17}) allows for high-precision astrometry of the Galactic center. This instrument has proven capable of an astrometric precision of $10-30 \mu \mathrm{as}$ during bright flares \citep{grav18b}. The observation of coherent circular motion of the centroid of emission during infrared flares lends new insight into the geometry and physical properties of matter in the accretion flow, as well as the spacetime near the black hole.  

In the past, efforts to predict near-infrared flares of the accretion onto Sgr A* have fallen into a few broad categories. The first of these involves detailed general relativistic magnetohydrodynamic simulations of a small volume around the black hole (e.g.,\ \citealt{dol12,dex13,cha15,bal16,res17}). These simulations allow for self-consistent modeling of the hydrodynamics of the accretion flow in the gravitational field of the black hole. However, the computational expense of these models preclude exploration of a large parameter space, making it difficult to constrain models based on observations. Simulations are usually able to produce only a limited number of orbits, making them sensitive to the initial conditions. Moreover, these models often fail to reproduce observed properties of Sgr~A* such as the wavelength-dependent variability or the polarization fraction or angle during flares. 

\citet{yua03} 
developed semi-analytic models of the spectral energy distribution and multi-wavelength variability of Sgr A*. These models successfully match the global observed properties of the black hole, but they do not explain the mechanism causing flares nor do they predict the distribution of matter near the black hole.  

Another class of recent models consist of Particle-in-Cell (PIC) simulations of a small region of the accretion flow (e.g.,\ \citealt{sir15,row17,row19}) . 
These models provide insight into the microphysical properties of the accretion flow, but they are computationally limited to very small regions. Predicting global observables, therefore, is not yet possible. 

The last class of models consists of semianalytic calculations of hotspots in orbits around the central black hole (e.g.,\ \citealt{bro05,pau05,bro06,mey06,tri07,ham09,zam10}). These models are computationally cheap enough to make it possible to explore a large parameter space of orbital configurations and black hole properties. However, they generally do not include any magnetohydrodynamic effects and, therefore, they cannot lend any insight into, for example, the mechanism that causes flares or the microphysical properties of the accretion flow. 

In the past, semianalytic hotspot models have been used to perform numerical calculations of photon geodesics through the curved spacetime near the black hole, but they have been limited to particle orbits that can be calculated analytically. In practice, this has restricted calculations to a small family of orbits--generally, circular orbits in the equatorial plane of the black hole. However, there is no a priori reason to expect the transiently heated, relativistic electrons responsible for infrared flares to be confined to the midplane or to circular orbits. 

In this paper, we describe the \textsc{nero} code, which generalizes the hotspot model to allow for arbitrary particle orbits in the vicinity of a Kerr black hole. Models with non-circular, unbound, or outflowing orbits are similarly allowed in our simulations. We discuss the methods for calculating photon (Section 2) and particle (Section 3) geodesics to generate synthetic flare observables incorporating all relativistic effects. We use the resulting model to analyze the GRAVITY data from 2018 (Section 4). We consider both circular orbits as in past works \citep{grav18b} and obtain new constraints on out-of-plane motion (Section 5), as well as the size of the emission region due to shear (Section 6). 

\section{Ray tracing}
The first step in calculating the appearance of the region near the black hole at the Galactic center is to find the paths followed by the photons as they travel towards a distant observer. We use the \textsc{geokerr} code \citep{dex09} to calculate null geodesics between the region around the black hole and a distant image plane. We assume a Kerr metric characterized by the line element 
\begin{eqnarray}
ds^2&=&-\left(1-\frac{2Mr}{\Sigma}\right)~dt^2
-\left(\frac{4Mar\sin^2\theta}{\Sigma}\right)~dtd\phi\nonumber\\
&&+\left(\frac{\Sigma}{\Delta}\right)~dr^2+\Sigma~d\theta^2\nonumber\\
&&+\left(r^2+a^2+\frac{2Ma^2r\sin^2\theta}{\Sigma}\right)\sin^2\theta~d\phi^2
\label{eq:kerr}
\end{eqnarray}
where $\Sigma = r^2 + a^2 \cos^2{\theta}$, $\Delta = r^2 - 2 M r + a^2$, $M$ is the mass of the black hole, and $a = J/ M^2$ is the dimensionless angular momentum. Here and throughout this paper, we use the convention $G = c = 1.$

There are two possible methods for finding particle trajectories in curved spacetimes. The more general approach is to directly solve the geodesic equations numerically. Alternatively, in certain astrophysically relevant metrics, the equations of motion can be reduced analytically to elliptical integrals (\citealt{cun73,rau94}). These integrals can then be calculated semi-analytically to find the particle trajectories.

The \textsc{geokerr} code adopts the latter approach to solve the geodesic equations in the Kerr metric \citep{dex09}. Starting from an image plane at the location of a distant observer, the code solves the geodesic equations backwards in time to find the trajectories of the photons originating near the black hole that terminate on the image plane. 

\section{Relativistic orbits}
The second component of the \textsc{nero} code is the calculation of the relativistic orbits of test particles near the event horizon of the black hole. We are particularly interested in calculating a broad range of orbits without constraining particles to be in circular orbits or in the equatorial plane of the black hole. 

Since we are interested in highly relativistic orbits close to the black hole, the usual Keplerian elements are a poor choice for characterizing orbits. Instead, we follow \citet{hug01} and identify each orbit by its energy $E$, $z$~angular momentum $L_z$, and Carter constant $Q$ in addition to choosing the spin $a$ of the black hole. While the procedure we use is capable of calculating general orbits with arbitrary inclination and eccentricity, we confine our initial analysis to circular orbits. In this case, we use the orbital radius $R$ as an orbital parameter along with $L_z$. We sample $R$ between the innermost stable circular orbit and an arbitrary large value (we choose $R = 14$ for the analysis below). 

In order to sample the range of allowed angular momenta, we first calculate the specific angular momentum of a prograde and retrograde orbit, respectively,
\begin{eqnarray}
L_z^{\rm pro} = \sqrt{R} \frac{1 - 2 a R^{-3} + a^2 R^{-4}}{\sqrt{1 - 3 R^{-2} + 2 a R^{-3}}}, \\
L_z^{\rm ret} = \sqrt{R} \frac{1 + 2 a R^{-3} + a^2 R^{-4}}{\sqrt{1 - 3 R^{-2} + 2 a R^{-3}}}.
\label{eq:Lz_pro_ret}
\end{eqnarray}
We can then create a family of orbits at a given radius by sampling $L_z$ evenly between these bounds to obtain orbits with a range of inclinations to the spin vector of the black hole.

The condition of circularity on the orbits introduces the constraints,
\begin{equation}
\frac{d\mathcal{R}}{d r} = \mathcal{R} = 0,
\end{equation}
where $\mathcal{R}$ is the effective potential in the radial direction.
\begin{equation}
\mathcal{R} = T^2 - \Delta \left(r^2 + \left(L - a E\right)^2 + Q\right)
\label{eq:Curly_R}
\end{equation}
These constraints allow us to find the specific energy and specific Carter constant for each value of $L_z$ and $R$. We can now find the energy and Carter constant of the orbit \citep{hug01}, 
\begin{eqnarray}
E = \left(a^2 L_z^2 (R - M) + R \Delta^2\right) / \nonumber\\
\bigg(a L_z M (R^2 - a^2) + \nonumber\\
\sqrt{R^5(R - 3M) + a^4R(R + M) + a^2R^2(L_z^2 - 2MR + 2R^2)}\bigg), \\
Q = \frac{\left[\left(a^2 + R^2 \right) E - a L_z \right]^2}{\Delta} - 
\left(R^2 - a^2 E^2 - 2 a E L_z + L_z^2\right).
\label{eq:E_Q}
\end{eqnarray}

Once we have calculated the constants of motion, we check to make sure the orbit is stable. The criterion for stability is 
\begin{eqnarray}
\frac{d^2\mathcal{R}}{d r^2} < 0. 
\label{eq:orb_stability}
\end{eqnarray}
Orbits in the region of the parameter space that does not satisfy equation~(\ref{eq:orb_stability}) are unstable and therefore astrophysically irrelevant.

For each orbit we wish to model, we have now calculated the constants of motion $E$, $L_z$, and $Q$. The equations of motion for a test particle following a timelike geodesic through the Kerr metric in terms of these constants are given by \citep{car68}:
\begin{eqnarray}
\label{eq:timelike_eom1}
\frac{dt}{d\tau} = \frac{1}{\Sigma}\left(-a\left(a E \sin^2{\theta} - L\right) + \frac{r^2 + a^2 T}{\Delta}\right)\\
\frac{dr}{d\tau} = \pm \frac{\sqrt{\mathcal{R}}}{\Sigma}\\
\label{eq:timelike_eom2}
\frac{d\theta}{d\tau} = \pm \frac{\sqrt{\mathcal{\Theta}}}{\Sigma}\\
\label{eq:timelike_eom3}
\frac{d\phi}{d\tau} = \frac{1}{\Sigma}\left(-a\left(E - \frac{L}{\sin^2{\theta}} + \frac{a T}{\Delta}\right)\right).
\label{eq:timelike_eom4}
\end{eqnarray}
Here, $\tau$ is the proper time, $\Theta$ is the effective potential in the angular direction,
\begin{equation}
\Theta = Q - \cos^2{\theta} \left(a^2\left(1 - E^2\right) + \frac{L^2}{\sin^2{\theta}}\right),
\label{eq:Th}
\end{equation} 
and $T$ is
\begin{equation}
T = E\left(r^2 + a^2\right) - L a.
\label{eq:big_T}
\end{equation}

We calculate the solution to these equations of motion using the publicly-available \textsc{ynogkm} code \citep{yan14}. This code adapts the \textsc{geokerr} algorithm to semi-analytically find particle trajectories in the Kerr-Newman metric for a given set of initial conditions. Since we are concerned with orbits about Sgr~A*, we set the charge of the black hole to zero, reducing the metric to the Kerr solution given in equation~\ref{eq:kerr}. 

The \textsc{ynogkm} code takes as input the initial position and velocity of a test particle. As an initial position, we arbitrarily choose a point in the equatorial plane ($\theta = \pi/2$) with a phase $\phi = \pi/2$, with $r$ set to the orbital radius $R$. The initial velocity in the Boyer-Lindquist frame can be calculated from the constants of motion:
\begin{eqnarray}
v^r_{\rm BL} = \pm \frac{\sqrt{\mathcal{R}}}{\gamma_p E \sqrt{\Sigma \Delta}}, \\
v^\theta_{\rm BL} = \pm \frac{1}{\gamma_p E \sqrt{\Sigma}}\sqrt{Q - \cos^2{\theta} \left(a^2\left(E^2 - 1\right) + \frac{\lambda^2}{\sin^2{\theta}}\right)},\\
v^\phi_{\rm BL} = \frac{\lambda}{\gamma_p E} \sqrt{\frac{\Sigma}{A \sin^2{\theta}}}, 
\label{eq:v_BL}
\end{eqnarray}
where
\begin{equation}
\gamma_p = \sqrt{\frac{A}{\Delta \Sigma}} \frac{1 - \lambda \omega}{E}
\label{eq:gamma_p}.
\end{equation}
We transform these velocities from Boyer-Lindquist coordinates into the locally non-rotating frame (LNRF, see \citealt{bar72}) for use by the \textsc{ynogkm} code:
\begin{eqnarray}
v^r_{\rm LNRF} = \sqrt{\frac{A_r}{\left(r^2 - 2 a + a^2\right)^2}} v^r_{\rm BL},\\
v^\theta_{\rm LNRF} = \sqrt{\frac{A_r}{\left(r^2 - 2 a + a^2\right)}} v^\theta_{\rm BL},\\
v^\phi_{\rm LNRF} = \sqrt{\frac{\left(1 - \mu^2\right)A_r^2}{\left(r^2 - 2 a + a^2\right) \left(r^2 + a^2 \mu^2 \right)^2}} \nonumber \\ 
\left(v^\phi_{\rm BL} - \frac{2 a r}{A_r}\right), 
\label{eq:v_coord}
\end{eqnarray}
where
\begin{equation}
A_r = r^4 + a^4 \mu^2 + a^2 r \left(2 + r - 2 \mu^2 + r \mu^2 \right).
\label{eq:Ar}
\end{equation}

The result of the \textsc{ynogkm} and \textsc{geokerr} codes is a set of position 4-vectors corresponding to the trajectories of the test particle as well as photons propagating to the observer. The intersections of these trajectories correspond to a point in spacetime at which a photon is emitted by the test particle that reaches the observer. To simplify the task of finding these intersections, we assign the hotspot a Gaussian emissivity profile with a size $s$ (set to a fiducial value of $s = 0.5M$):
\begin{equation}
    \epsilon \propto e^{\frac{-\bf{x}^2}{2 s^2}},
    \label{eq:emissivity_density}
\end{equation}
where $\bf{x}$ is the distance from the center of the hotspot.

Given the positions and velocities of a particle as well as the constants of motion of the photon trajectory, we can calculate the Doppler shift of the intersection as 
\begin{equation}
D = \mathbf{u} \cdot \mathbf{v}
\label{eq:D_u_v}
\end{equation}
In terms of the metric elements defined above, we write this Doppler shift as
\begin{equation}
D = \gamma \sqrt{\frac{A}{\Delta \rho^2}} \left( 1 - \lambda \Omega \pm v^r_{\rm LNRF}\sqrt{\frac{\mathcal{R}}{A}} \pm v^\theta_{\rm LNRF}\sqrt{\frac{\Delta \Theta}{A}}\right),
\label{eq:D_metric}
\end{equation}
where $\gamma = 1/\sqrt{1 - v^2/c^2}$ is the usual relativistic factor, 

With our assumption that the emission is optically thin, we can now find the flux observed at the image plane by integrating the emissivity scaled by the Doppler term backwards along the photon trajectories calculated from the \textsc{geokerr} code:
\begin{equation}
    F = \int D^2 \epsilon ds,
    \label{eq:flux_los_integral}
\end{equation}
The time of arrival of the photons at the image plane is the sum of orbital time and the light-travel time from the hotspot to the observer. 

\begin{figure*}
\includegraphics[width=7in]{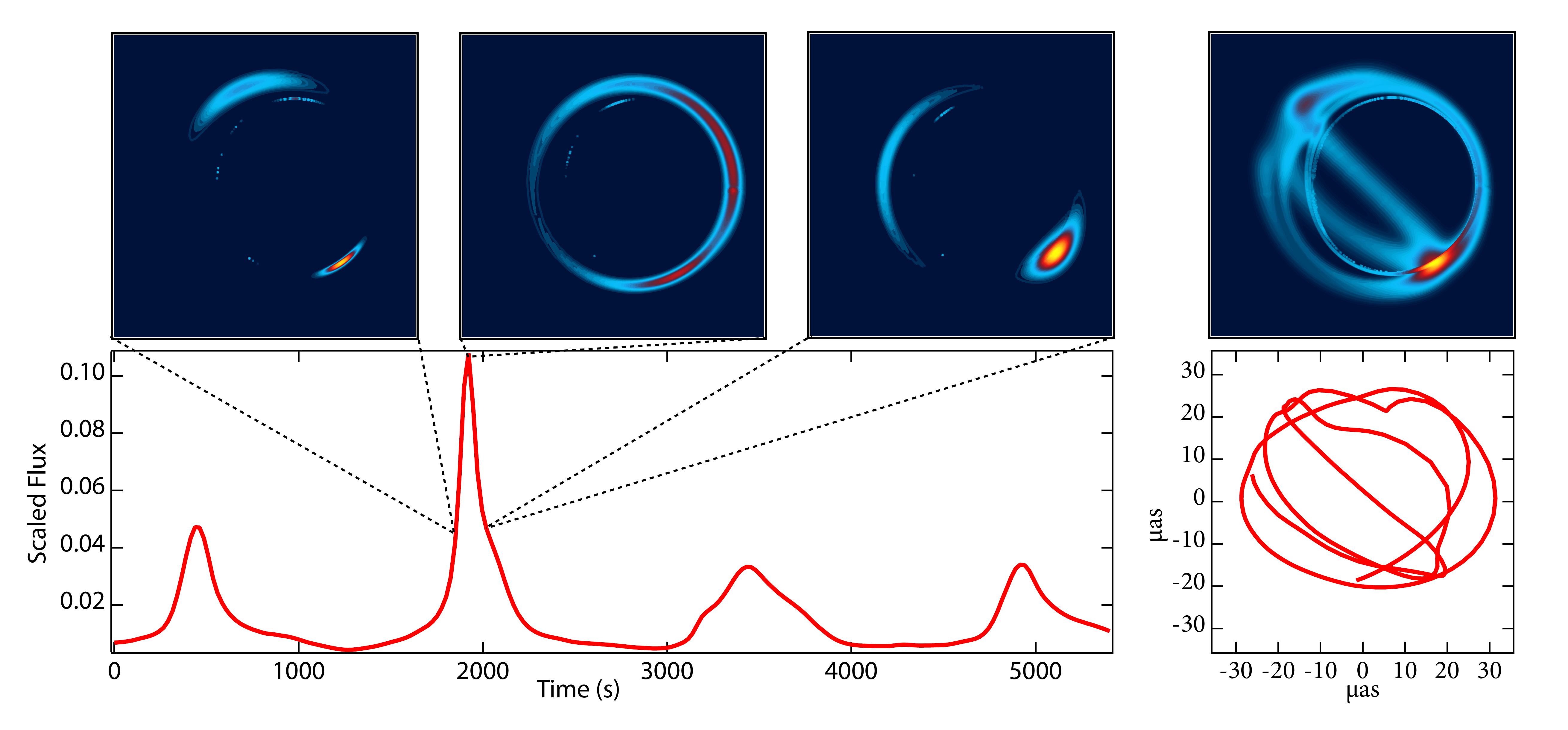}
\caption{Sample image of a hotspot at different points in the orbit. Top row shows the image of the hotspot before, during, and after a strong lensing event. Top-right plot shows the image averaged over the whole orbit. The lower left panel shows the lightcurve produced over several orbits, while the lower right panel shows the motion of the centroid of emission. This figure was produced with a hotspot in a circular orbit at $R = 5M$ around a black hole with spin $a = 0.9$. The spin axis of the black hole is inclined $i = 15^\circ$ from the observer's line of sight and the relative inclination between the spin axis and the orbital plane is $\theta = 50.54^\circ$.}
\label{fig:panel}
\end{figure*}

The end result of our simulation is a series of snapshots at increasing observer timesteps of the region immediately around the central black hole. For each image, we can then calculate a centroid position,
\begin{equation}
{\bf c} = \iint {\bf x} F(x,y) dx dy, 
\label{eq:cen_pos}
\end{equation}
and a flux,
\begin{equation}
F = \iint F(x,y) dx dy, 
\label{eq:flux_int}
\end{equation}

Figure~\ref{fig:panel} shows an example series of images of a hotspot passing behind the black hole. As the spot moves behind the black hole, the photon trajectories are strongly lensed in the direction of the observer, resulting in the corresponding peak in the lightcurve. Simultaneously, the centroid of the emission moves towards the location of the black hole.

\section{Orbit fitting}
The GRAVITY instrument has found evidence of orbital motion around Sgr A*. Here we describe the process used to fit the observed motion with the \textsc{nero} code to obtain constraints on the properties of the emitting region \citep{grav18b}.

We construct a grid of model centroid motions across our parameter space. For the purposes of this work, we limit 
ourselves to fitting only for the orbital radius and inclination to the observer. Therefore, we calculate a range of models between $R = 4$ and $R = 14$ in increments of $\Delta R = 0.5$ across all inclinations ($1^\circ < i < 180^\circ$, $\Delta i = 5$). Inclinations between $0^\circ$ and $90^\circ$ correspond to counterclockwise orbits on the sky, while inclinations above $90^\circ$ correspond to clockwise orbits. At each point in this grid, we further vary the position angle of the orbit on the sky, the orbital phase at the start of the observation, and the x- and y-position of the central mass. For each parameter combination, we calculate the $\chi^2$ difference between the model and the astrometric data. The location of the minimum $\chi^2$ value determines the best fit to the observation.

\begin{figure*}
    \centering
    \includegraphics[width=7.5in]{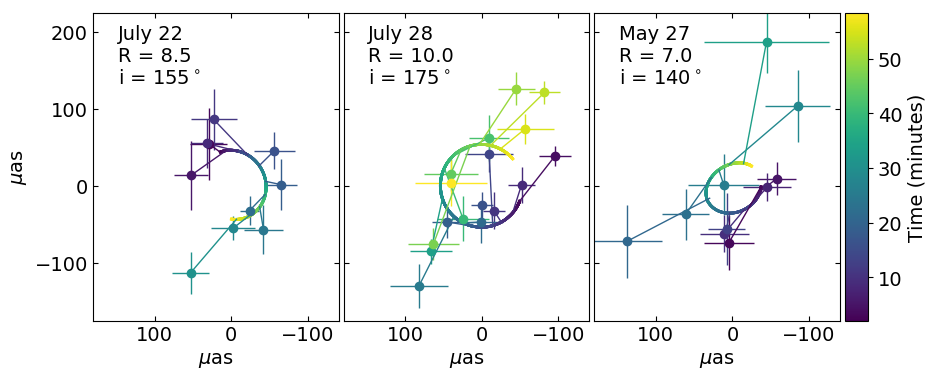}
    \caption{Data and \textsc{nero} fits to the flares from May 27, July 22, and July 28, 2018. Data points correspond to the centroid position observed by GRAVITY over the course of the flare. The solid lines show the orbits that best fits each of three flares. Colors indicate the time since the start of the flare.}
    \label{fig:Data_Fit_Panel}
\end{figure*}

Figure~\ref{fig:Data_Fit_Panel} shows the best-fit models for each of the three flares observed during the summer of 2018. The reduced $\chi^2$ for each fit is 1.6, 4.5, and 2.3 for July 22, July 28, and May 27, respectively. In each case, the best-fit model lies mostly or entirely within the astrometric data points. This is due to our assumption that the orbiting hotspot follows a geodesic of motion, meaning that its orbital speed is determined by its distance from the black hole. At face value, the data from all three nights show motion that is faster than the expected orbital speed or equivalently, a radius that is larger than expected.

One effect that partially accounts for this tension is related to the scatter in the observed centroid points. Since the distance from the center of the orbit is a positive definite quantity, any scatter in the measurements will tend to increase the observed radius by a factor of $\sqrt{1 + \sigma^2/R^2}$. In the case of an orbit at $R = 7M$ observed with $30 \mu$as error bars, this effect implies a $10 \mu$as bias in the observed radius. Other possible explanations are that the motion of the hotspot is not in the plane of the black hole but rather moving outward towards the observer (see Section~6 below) or that other forces, such as a pressure gradient or the magnetic field, have a significant effect on the motion of the hotspot. 

\begin{figure*}
    \centering
    \includegraphics[width=6.5in]{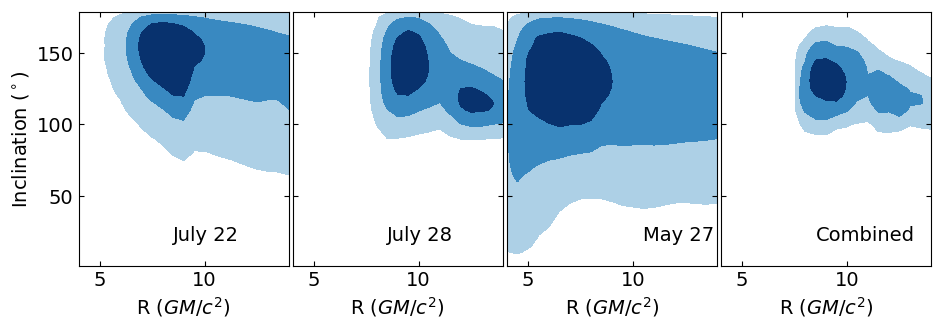}
    \caption{Confidence contours corresponding to $1-$, $2-$, and $3\sigma$ limits on the orbital radius and inclination for all three flares. The last panel shows the combined constraints, assuming the radius, inclination, and position angle are the same for all three flares.}
    \label{fig:3_Contours_Panel} 
\end{figure*}

We marginalize over the various parameters to find the 1-, 2-, and 3-$\sigma$ allowed regions, corresponding to the area containing $39\%$, $86\%$, and $98.8\%$ confidence regions. Given the large values of the $\chi^2$ of the best-fit models, we scale the uncertainties of the model parameters such that the reduced $\chi^2$ of the best-fit model is 1. The first three panels of Figure~\ref{fig:3_Contours_Panel} show the constraints in the R-i plane for all three flares. The clockwise motion during the flares constrains the inclination of the orbit to be between $\sim120^\circ - 175^\circ$ for all but the May 27 flare. In the latter case, the best-fit inclination is also in this range ($140^\circ$) but the allowed region extends over a larger fraction of the parameter space due to the poor astrometric precision of the data. Unless otherwise noted, we include a $\sin{i}$ prior on the inclination of the orbit to account for the geometric bias favoring edge-on orientations. We find that the astrometric fit favors models with moderate to high inclinations and radii between $7$ and $12M$. The allowed parameter spaces of all three flares overlap, indicating that a single model is consistent with all the flares.

\begin{figure}
    \includegraphics[width=3.5in]{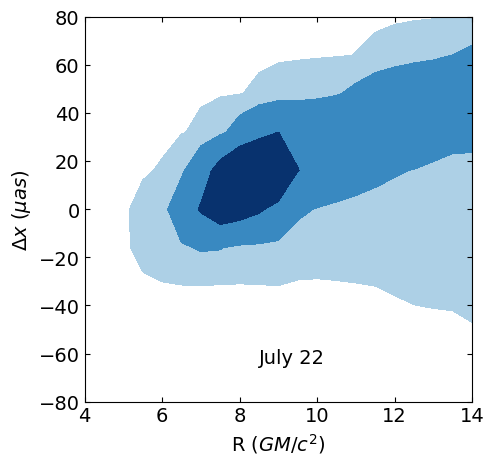}
    \caption{Confidence contours showing the allowed values for the orbital radius and the x-offset of the center of the black hole for the July 22, 2018 flare. An offset of zero corresponds to the black hole lying at the mean location of emission averaged over the entire flare.}
    \label{fig:R_dx}
\end{figure}

The contours shown in Figure~\ref{fig:3_Contours_Panel} are slightly less constraining than those found in \citet{grav18b}. This difference is due to the additional degrees of freedom allowed in the fit corresponding to the uncertainty of the position of the black hole. In previous fits, the center of the orbit was placed at the average of all the astrometric points over the course of the flare. Figure~\ref{fig:R_dx} illustrates the degeneracy between the x-position of the central mass and the radius of the orbit. This degeneracy indicates that the flare astrometry allows for models in which the flare represents only a small fraction of an orbit with a large radius, where the black hole is correspondingly offset from the center of emission. However, these models can be ruled out based on our knowledge of the position of the center of mass from the measured orbits of the S-stars. In particular, the orbit of the star S2 constrains the position of the center of mass to within $\sim 50 \mu$as \citep{grav18a}.

The fits presented here only make use of the measurements of the centroid position of the emission throughout the flare. Two additional sources of information are available that can further constrain the parameters of the fit. The first is the absence of clear Doppler beaming during a portion of the flare. For orbits that are close to edge-on to the observer, the relativistic speed at which the hotspot moves should lead to significant brightening on the approaching side and dimming on the receding side. These effects are, of course, convolved with the intrinsic light curve of the flare, making it difficult to place precise constraints on the inclination. However, orbits with inclinations between $40^\circ$ and $140^\circ$ would lead to Doppler beaming comparable to the overall brightening of Sgr A* during the flare, making these scenarios unlikely \citep{grav18b}.

\begin{figure}
    \includegraphics[width=3.5in]{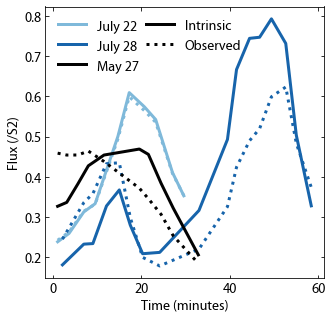}
    \caption{Observed and inferred intrinsic lightcurves for the three GRAVITY flares. The flux is measured relative to the nearby star S2. Each dotted line corresponds to the lightcurve during the respective flare as observed by GRAVITY. The solid line shows the intrinsic lightcurve once all relativistic effects predicted from the best-fit orbital model have been remove.}
    \label{fig:inferred_lightcurves}
\end{figure}

Since we do not know the underlying mechanism for generating flares, it is impossible for us to construct light curves that can be compared directly to the data. However, the \textsc{nero} code does calculate the Doppler boosting, lensing, and other relativistic effects of the motion of the hotspot on the observed flux. By combining these predictions with the observed lightcurves, we can find the intrinsic flare lightcurve as generated at the source. Figure~\ref{fig:inferred_lightcurves} shows both the observed and inferred intrinsic lightcurves for each of the three flares discussed in this paper.   

A further constraint arises from the observed change in the polarization angle during the course of the flare. The polarization angle changes on a timescale similar to the orbital timescale at the radii favored by the astrometric data. A detailed analysis of the constraints arising from the polarization will be addressed in a forthcoming paper (Jim\'enez-Rosales et al.\ in prep.).  

If we make the assumption that all of the flares share the same parameters, we can obtain a combined constraint for all of the flares. In order to calculate the combined constraint, we allow the orbital phase at the start of each observation as well as the x- and y-position of the central mass to vary between observations. We hold the radius, inclination, and position angle of the model constant across all three observations. The last panel of Figure~\ref{fig:3_Contours_Panel} shows the constraint on the $R$-$i$ parameter space from all three flares combined. The preferred radius is between $8$ and $10M$, and the inclination lies between $i = 120^\circ$ and $i = 150^\circ$.

\begin{figure}
    \centering
    \includegraphics[width=3.5in]{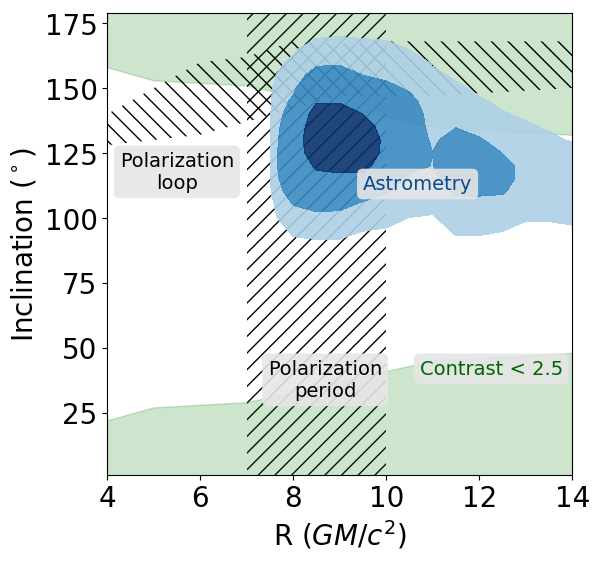}
    \caption{Combined constraints together with information from the polarization and the Doppler beaming. The blue astrometric constraints are as shown in Figure~\ref{fig:3_Contours_Panel}. The green shaded region corresponds to models where the maximum contrast of the lightcurve is less than $2.5$. The hatched polarization period constraint arises from the assumption that the change in polarization angle during the flare is caused by the motion of the hotspot through the magnetic field. The polarization loop constraint is based on a model with a vertical magnetic field and the requirement that the polarization angle matches the behavior observed during the flare. }
    \label{fig:B3}
\end{figure}

Figure~\ref{fig:B3} shows the best-fit region from the astrometry combined with the constraints from the lightcurve and the polarization. The contrast constraint is based on a conservative upper limit on the amount of Doppler boosting that is consistent with the observed lightcurve during the flare. The two hatched regions correspond to the radius range determined from the rate of change of the polarization angle and the behavior of the polarization angle in the Q-U plane. We note that this constraint is based on the assumption that the magnetic field is vertically oriented in the emitting region of the accretion flow (see \citealt{grav18b} as well as Jim\'enez-Rosales et al.,\ in prep.\ for details). 

\begin{figure}
    \centering
    \includegraphics[width=3.5in]{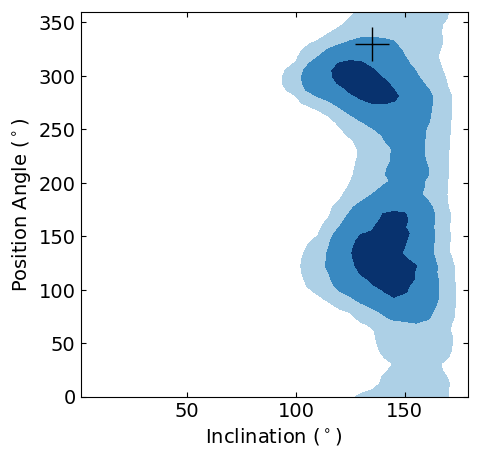}
    \caption{Combined constraints on the orientation of the hotspot orbit from all three flares. The black marker indicates the approximate orientation of the clockwise disk of stars.}
    \label{fig:i_phi0}
\end{figure}

We also show the constraints on the orientation of the orbiting hotspots in Figure~\ref{fig:i_phi0}. Interestingly, the allowed region of of orbital orientations is broadly consistent with those of the stars in the clockwise stellar disk, as marked by the black cross in the figure (\citealt{gil17}, see also \citealt{bar09,yel14}). This clockwise disk consists of O-stars with significant stellar winds \citep{pau06}. These winds have been conjectured as a possible source of matter to fuel the accretion onto the central black hole. If this is the case, the accretion flow is expected to roughly align with the stars in the clockwise disk (\citealt{res18,cal19}). While our constraints cannot definitively confirm this picture, improved constraints from future observations may be able to pinpoint whether the accreting material originates in the winds from massive stars in the clockwise disk. 

\section{Orbital shear}

If the circular centroid motion during infrared flares is indeed due to the orbit of a hotspot, the hotspot is expected to shear over the course of its orbit due the different orbital velocity at different radii. Although the shape of the hotspot is unresolved, this shearing would manifest itself as an inward spiral of the centroid over the course of the flare. By incorporating the orbital shear into our model of the centroid motion and comparing this prediction to the observed astrometry, we can, in principle, obtain an upper limit on the size of the emitting region.

We allow for the possibility of shearing by modifying our model to account for the differential orbital velocity at different points in the emitting region. We define a parameter $s$ corresponding to the radial extent of the hotspot. For a given orbital radius $R$, we calculate a series of particle trajectories with starting positions ranging from $R - s$ to $R + s$. We then sum the emission from a Gaussian hotspot on each of these geodesics to find the total predicted lightcurve and centroid track. 

\begin{figure}
    \includegraphics[width = 3.5in]{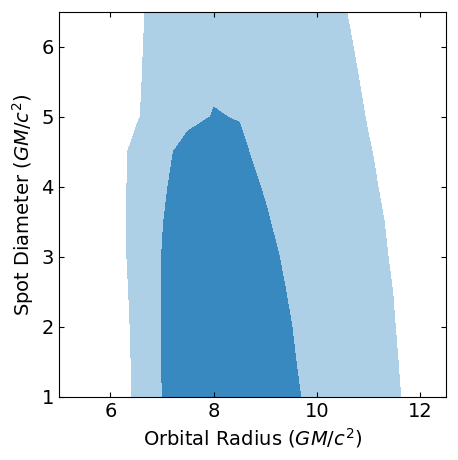}
    \caption{1- and 2-$\sigma$ constraints on the orbital radius and spot diameter for the July 22 flare.}
    \label{fig:Shear_Contour_Panel}
\end{figure}

We use the same procedure described above to fit the resulting sheared-spot models to the data. In this case, we fix the inclination and restrict ourselves to orbital radii between $5M$ and $12.5M$. We let the diameter of the emitting region vary between $1.0M$ and $6.5M$. Given the poor astrometric precision of the flare observations on May 27 and July 28, we have restricted this analysis to the July 22 flare. Figure~\ref{fig:Shear_Contour_Panel} shows contours of constant $\chi^2$ corresponding to 1-, and 2$\sigma$ limits on the orbital radius and spot extent. As the size of the spot increases, the quality of the fit decreases. Spots larger than $5M$ in diameter can be ruled out at the 1-$\sigma$ level, although larger spots are still within the 2-$\sigma$ limits from the current observations. 

\section{Out-of-plane motion}

Although the detection of circular motion during flares would seem to favor the interpretation that flares originate in accreting material, it is, in principle, also consistent with material launched at the base of a jet. This material would retain some angular momentum from the accretion flow and therefore spiral outward from the jet-launching region.  

\begin{figure}
    \centering
    \includegraphics[width = 3.5in]{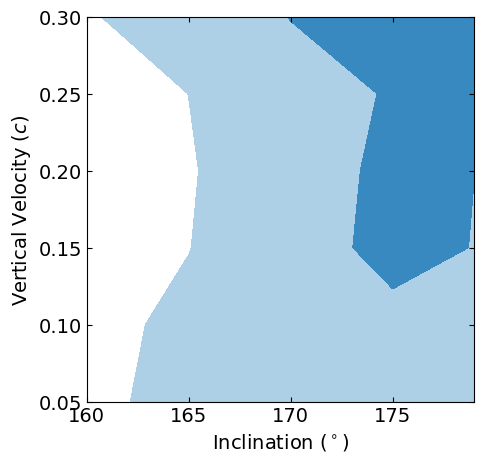}
    \caption{Constant $\chi^2$ contours corresponding the the 1-, and 2$\sigma$ regions of the vertical velocity and inclination for the July 22 flare. We note that in this case we use a flat prior for the inclination of the orbit.}
    \label{fig:Contour_vz}
\end{figure}

To investigate this possibility, we simulate the motion of material launched vertically from an accretion flow. In particular, we modify our existing models of orbital motion by adding a constant vertical component $v_z$ to the velocity of the hotspot. We then renormalize the four-velocity of the hotspot to conserve the relation $u \cdot u $ = -1. The result is a hotspot on a helical track originating in the equatorial plane. Since the volume of the region we simulate is limited for computational reasons, we have restricted ourselves to scenarios where $v_z < 0.3c$ to prevent the hotspot from leaving the bounds of our simulation during a single orbit. 

We again use the procedure described above to fit the parameters of the model to the July 22 flare. If the axis of a jet is significantly inclined with respect to the observer, the hotspot should display a systematic linear motion over time. The relatively circular motion measured by GRAVITY rules out this scenario. Therefore, we restrict the inclination to be greater than $160^\circ$. Given the face-on inclination inherently required by this scenario, we do not weight the probability distribution by $\sin{i}$ as above. We allow the vertical velocity to vary between $0.05c$ and $0.3c$. 

Figure~\ref{fig:Contour_vz} shows the allowed region of the parameter space for the July 22 flare. As expected, the combination of high inclination and large vertical velocity has been ruled out. For a sufficiently low inclination, however, the primary effect of the vertical velocity is to change the observed orbital period due to the propagation time of the photons and the decreasing distance to the hotspot. For face-on orbits, the observed centroid motion is consistent with a significant vertical velocity. In fact, adding a vertical velocity component marginally improves the quality of the fit, with the best-fit parameters being $R = 9.0M$, $i = 175^\circ$, $v_z = 0.15c$.  

\begin{figure*}
\includegraphics[width=6.5in]{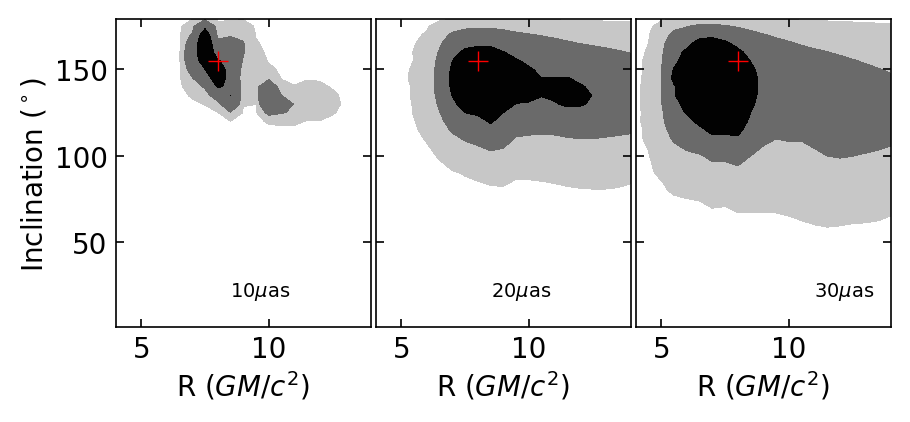}
\caption{Constraints from simulated data  with improved error bars. In each case, the underlying model is based on a hotspot orbiting at $R = 8.0M$ with an inclination of $i = 155^\circ$, as marked by the red cross in each panel. The astrometric error is $10 \mu$as, $20 \mu$as, and $30 \mu$as, respectively.}
\label{fig:fake_contour_panel}
\end{figure*}

\section{Prospects for improved constraints}

There are two possible avenues for improving the constraints on the flare properties from astrometric measurement. The first consists of continued observations of flares at the current level of precision. As shown in the last panel of Figure~\ref{fig:3_Contours_Panel}, co-adding several flares reduces the statistical error on the inferred parameters. Additional flare observations will further tighten the constraints. However, these improvements are contingent on the assumption that some parameters of the flares (i.e.,\ the radius, inclination, and position angle) are consistent between multiple flares. 

The second avenue for improved constraints stems from increasing the astrometric precision of measurements of individual flares. To simulate this scenario, we have generated simulated data sets at various levels of precision and used the \textsc{nero} code to find the best-fit model and resulting constraints on the flare properties. Figure~\ref{fig:fake_contour_panel} shows the confidence contours for three such simulations. In each case, the underlying model is a hotspot orbiting at a radius of $R = 8M$ and an inclination of $i = 155^\circ$. Each simulated data set consists of ten data points chosen at time intervals corresponding to the observed times of the July 22nd flare. The x- and y-data points are drawn from a normal distribution centered on the model value with a standard deviation corresponding to an error of 10, 20, and 30$\mu$as, respectively.

We then apply the same fitting procedure to our simulated data points as we use on the real data. The constraints from these fits are shown in Figure~\ref{fig:fake_contour_panel}. For small error bars, the radius and inclination are tightly constrained. We correctly recover the parameters of the underlying simulation, as marked in the figure. For a 30$\mu$as error, which most closely corresponds to the current GRAVITY precision, the simulated data are consistent with radii between $8M$ and $14M$. The inclination is only loosely constrained.

\section{Summary and discussion}
The \textsc{nero} code combines the ability to numerically calculate relativistic orbits near a black hole with efficient ray-tracing between the vicinity of the black hole and a distant observer. The result is a hotspot model for fitting both astrometric and light-curve data with the ability to account for the full general relativistic treatment of the spacetime near the black hole. It also takes into account a variety of physical scenarios, including hotspot orbits that are inclined to the spin axis, eccentric orbits, out-of-plane motion, and orbital shear from hotspots with a significant radial extent.

We apply this model to the three flares observed during the summer of 2018 with the GRAVITY instrument at the VLT. We find that all three flares are well fit by hotspots on circular orbits with radii between $7$ and $10M$ and inclinations greater than $\sim 90^\circ$. A single model with $R = 9M$ is consistent with all three observed flares at a range of inclinations. 

While the observed centroid motions are consistent with circular orbital motion, the size of the observed motion on the sky seems to be systematically larger than what has been predicted by our models. 
Equivalently, the period of the motion is smaller than predicted by the observed size of the centroid track. 
One possible explanation for this discrepancy is that the hotspot is moving towards the observer at a significant fraction of the speed of light, causing the observed orbit to appear faster than it really is. 
We find that models with an additional vertical velocity component are consistent with the observations, but they are not significantly preferred over orbits in the equatorial plane. 

We also investigate the effects of orbital shear on the predicted motion of the centroid. For a hotspot with a size comparable to its orbital radius, the differential velocity at different radii should lead to a centroid track that spirals inward over the course of the orbit. We find that smaller hotspots provide a better fit to the observations, but hotspots with diameters as large as $6M$ are allowed by the current astrometric precision.  

Additional constraints on the viability of a model of orbital motion comes from the observed light curve and polarization of the flare. A hotspot orbiting at a high inclination to the observer is expected to show significant variation in brightness due to the difference in Doppler beaming between the approaching and receding parts of the orbit. However, this constraint is significantly complicated by the unknown emission mechanism of infrared flares, which leads to a large uncertainty in the expected light curve. In addition to astrometric motion during infrared flares, the polarization angle of the emission also varies on a timescale consistent with the orbital period of a hotspot. Using these observations to constrain the orbit requires a model of the magnetic field near the black hole and will be presented in a future work. 

At the current level of precision, astrometric measurements during infrared flares provide strong evidence that matter is in orbital motion around the central black hole. The \textsc{nero} code allows for relativistic orbits to be fit to the astrometric data in order to constrain the parameters of the orbits. The combination of multiple flare observations can reduce the uncertainties on the parameters of the orbiting material.
Further observations of greater precision will be able to improve these constraints even further.

\bibliographystyle{aa}
\bibliography{bibliography}

\end{document}